\documentclass[preprint2]{aastex}
\usepackage{graphicx}
\usepackage{times}
\usepackage{bm}
\usepackage{verbatim}
\usepackage{fullpage}

\newcommand{\bK}{ \mbox{\boldmath$K$} }

\newcommand{\bA}{ \bm{\mbox{A}} }
\newcommand{\bb}{ \bm{\mbox{b}} }

\newcommand{\bw}{ \mbox{\boldmath$w$} }

\newcommand{\bu}{ \mbox{\boldmath$u$} }

\newcommand{\bxhat}{{\mbox{\boldmath$\hat{x}$}}}
\newcommand{\br}{{\mbox{\boldmath$r$}}}

\newcommand{\bcK}{ \mbox{\boldmath$\cal K$} }
\newcommand{\bcT}{ \mbox{\boldmath$\cal T$} }
\newcommand{\cK}{ \mbox{$\cal K$} }
\newcommand{\cT}{ \mbox{$\cal T$} }

\begin{document}

\title{A procedure for the inversion of f-mode travel times for solar flows}

\author{J. Jackiewicz}
\affil{Max-Planck-Institut f\"{u}r Sonnensystemforschung, 37191 Katlenburg-Lindau, Germany}
\author{L. Gizon}
\affil{Max-Planck-Institut f\"{u}r Sonnensystemforschung, 37191 Katlenburg-Lindau, Germany}
\author{A.~C. Birch}
\affil{CoRA Division, NWRA, 3380 Mitchell Lane, Boulder, CO 80301, USA}
\author{M.~J. Thompson}
\affil{Department of Applied Mathematics, University of Sheffield, Hounsfield Road, S3 7RH Sheffield, United Kingdom}





\keywords{Helioseismology, Sun: convection}

\begin{abstract}
We perform a two-dimensional inversion of f-mode travel times to determine
near-surface solar flows. The inversion is based on optimally localized
averaging of travel times. We use finite-wavelength travel-time sensitivity
functions and a realistic model of the data errors. We find that it is
possible to obtain a spatial resolution of 2~Mm. The error in the resulting
flow estimate ultimately depends on the observation time and the number of
travel distances used in the inversion.
\end{abstract}


\section{Introduction}

In time-distance helioseismology \citep{duvall1993}, one measures the wave travel time from one point on the surface of the Sun to another. Local flows interact with the waves and affect the travel times. The travel-time measurements thus contain information about the internal flows.

One major goal is to infer the near-surface flows at the highest spatial and temporal resolution possible, in order, for instance, to study supergranulation in detail. Previous studies have shown that this is indeed possible using time-distance helioseismology \citep{gizon2000, duvall2000}. To achieve the desired accuracy and resolution, a consistent inversion must be carried out, which involves two main steps. The first step, called the forward problem, is to model the wave-flow interaction to obtain sensitivity kernels that relate the travel times to small-amplitude flows. The second step is to use the kernels and the travel-time measurements, together with noise estimates, to infer the flows.

Here we invert f-mode travel times for two - dimensional, spatially-varying, horizontal flows. We use the sensitivity kernels of \citet{jackiewicz2006}, which are computed in the first Born approximation. These kernels are two dimensional as they are averaged over the depth over which the f-mode kinetic energy is significant. We expect the inversion to return horizontal flows at an average depth of 1~Mm below the surface. We note that the kernels are sensitive to flows at horizontal scales as small as half the f-mode wavelength of about 5~Mm at 3~mHz.

The inversion scheme adopted in this study is the Subtractive Optimally Localized Averaging (SOLA) method \citep{pijpers1992,pijpers1994}, which is commonly used in helioseismic inversions but has not yet been demonstrated for the time-distance technique. To keep the presentation simple, we formulate the two-dimensional inversion for only one travel distance of $4.96$~Mm. The method can be generalized for many distances.

In Section~\ref{sec:problem} we discuss briefly the inputs to the inversion, in Section~\ref{inversion} we formulate the two-dimensional SOLA inversion scheme, and in Section~\ref{results} the results of the inversion are shown. We conclude in Section~\ref{discussion}.

\section{Inputs to the inversion}
\label{sec:problem}

\begin{figure*}[t]
   \includegraphics[width=7.in, clip]{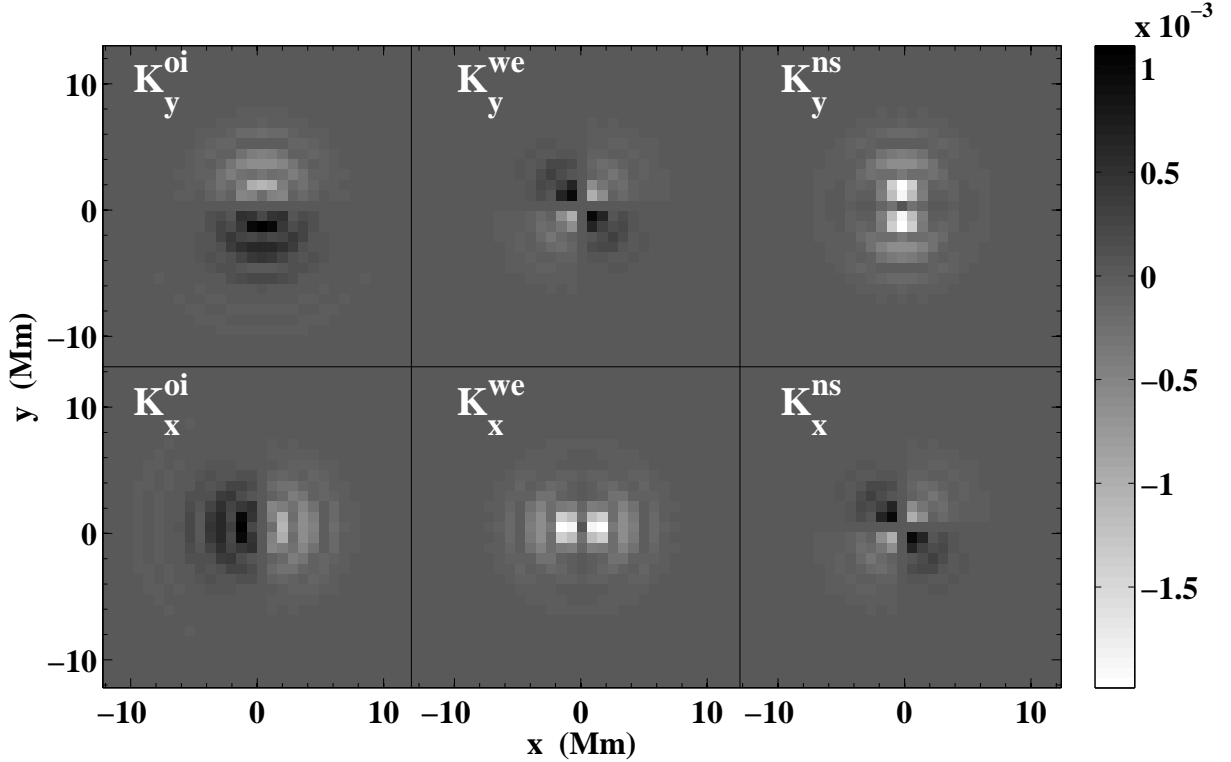}
\centering
   \caption{Travel-time sensitivity kernels used in this study, $\bK^\alpha= (K_x^\alpha, K_y^\alpha)$, where $\alpha$ stands for `oi', `we', `ns'. The units of the kernels are s~(m/s)$^{-1}$.  The grid spacing is ${\rm d}x={\rm d}y=0.826$~Mm. The center of each annulus lies at $\br=(0,0)$. The spatial sum of $K_x^{\rm we}$ and $K_y^{\rm ns}$ is $-0.043$~s~(m/s)$^{-1}$. The sums of all other kernels are zero.}
   \label{fig:ave_kerns}
\end{figure*}

We work in a Cartesian coordinate system ($x,y$) representing a small area of the solar surface. The $x$ coordinate is measured in the pro-grade direction (toward the west limb) and the $y$ coordinate is measured toward the north pole.  The quantities needed for the inversion are the travel-time measurements, the sensitivity kernels, and the noise-covariance matrix. We briefly outline these below.

We choose to consider three different types of f-mode travel-time measurements. These are the `oi', `we', and `ns' travel times, which refer to the out minus in, west minus east, and north minus south travel-time differences, respectively. These measurements are very similar to the spatial averages of travel times over quadrants that have been introduced by \citet{duvall1997}. A difference is that we are using the travel-time measurement technique described by \citet{gizon2004}. The `oi' travel times are measured between a center point and a concentric annulus of radius $4.96$~Mm; they are given by the difference in travel time between waves that propagate outward from and inward toward the center point. The `we' travel times are obtained by measuring the travel time using the wave signal at the center point and the signal averaged over the annulus with a weighting of $\cos\theta$, where $\theta$ is the angle between the $x$ direction and each point on the annulus. Thus, the `we' travel time is mostly sensitive to flows in the $x$ direction. Similarly, the `ns' travel time is obtained using a weighting of $\sin\theta$ to give sensitivity to flows in the $y$ direction.

We use the short notation $\tau^\alpha(\br_i)$ to denote the travel times; the superscript $\alpha$ is one of `oi', `we' or `ns'. The horizontal vector $\br_i$ gives the center position of the $i$-th annulus and is defined on a grid with spacing ${\rm d}x={\rm d}y=0.826$~Mm in both the $x$ and $y$ directions (twice the MDI high-resolution spatial sampling). Let us denote the horizontal flow by $\bu = (u_x, u_y)$. When the flow is weak, the relationship between the travel times and $\bu$ can be written as
\begin{equation}
\label{tau-u}
\tau^\alpha(\br_i) = \sum_j \bK^\alpha(\br_j-\br_i)\cdot\bu(\br_j) + n^\alpha(\br_i),
\end{equation}
where the two-dimensional vector-value functions $\bK^\alpha=(K_x^\alpha,K_y^\alpha)$ are the sensitivity kernels. The quantity $n^{\alpha}$ denotes the random error associated with the measurement of $\tau^\alpha$. The sum runs over all spatial pixels.  Note that to simplify the notation, we have discretized the flow.

The f-mode sensitivity kernels $\bK^\alpha$ are $\alpha$-averages of the point-to-point Born sensitivity kernels presented in \citet{jackiewicz2006}. The six kernels used in this study are shown in Figure~\ref{fig:ave_kerns}.

\begin{figure}[t]
\begin{center}\includegraphics[width=3.7 in, clip]{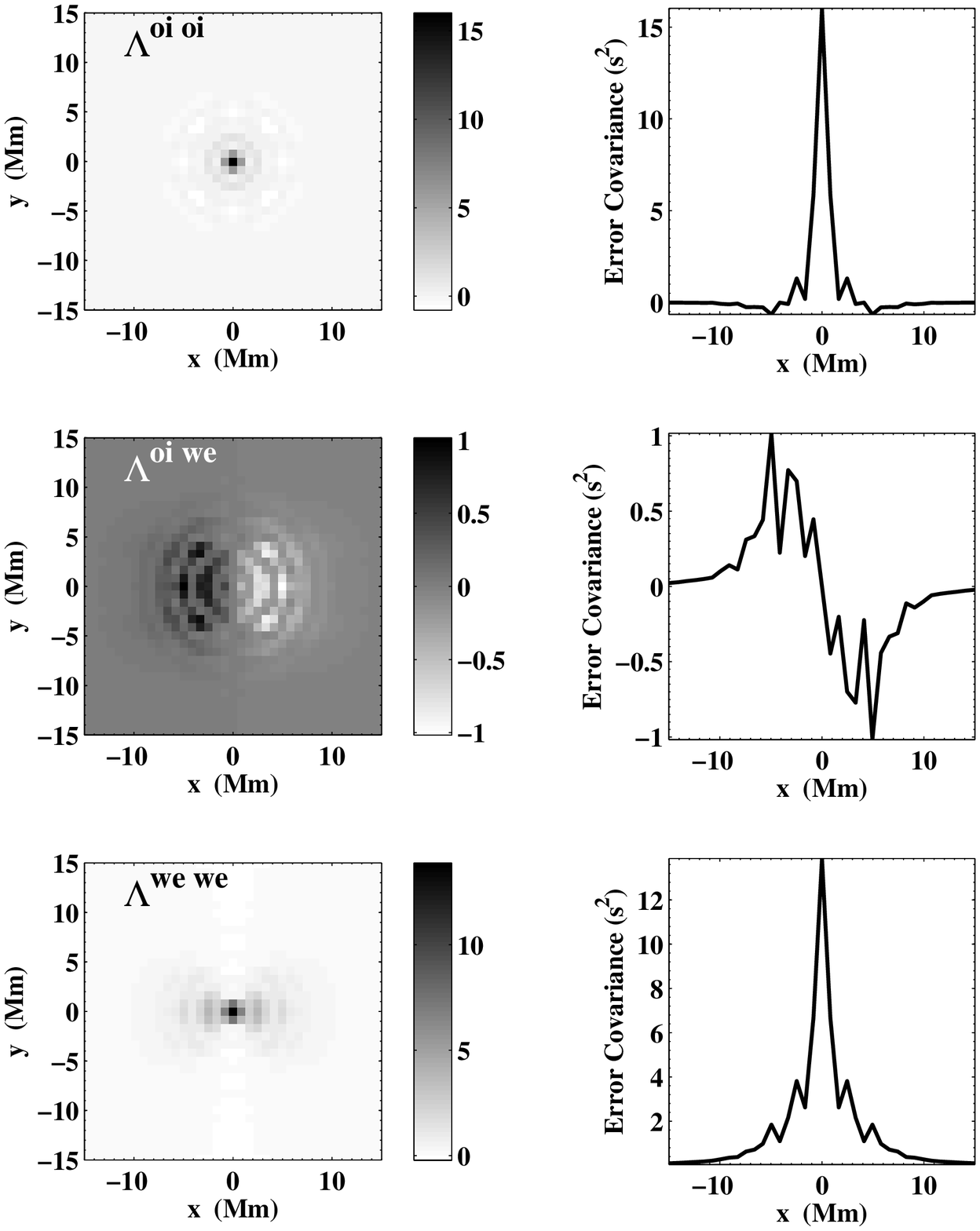} \end{center}
\caption{Example calculations of the noise covariance, $\Lambda^{\alpha\beta}$. The left column shows the maps $\Lambda^{{\rm oi}\,{\rm oi}}(\br)$ (top), $\Lambda^{{\rm oi}\,{\rm we}}(\br)$ (middle), and $\Lambda^{{\rm we}\,{\rm we}}(\br)$ (bottom) as functions of $\br=(x,y)$. The units are s$^2$ and the observation time is $T=12$~hr. The right panels are cuts through the corresponding maps at $y=0$. }
   \label{fig:error_cov}
\end{figure}

The statistical properties of the noise in travel times were discussed in detail by \citet{gizon2004}. They are specified by the covariance matrix $\Lambda$,
\begin{equation}
\Lambda^{\alpha\beta}(\br_i-\br_j)={\rm Cov}\left[n^{\alpha}(\br_i),n^{\beta}(\br_j)\right] .
\end{equation}
We compute $\Lambda$ according to equation~(28) of \cite{gizon2004} and using the model f-mode power spectrum from \citet{jackiewicz2006}.  The covariance matrix elements scale with the observation time $T$ as $1/T$. To speed up the computation, the full matrix $\Lambda(\br)$ can be constructed using a subset of $\br$ values and the symmetries of the problem. Figure~\ref{fig:error_cov} shows example calculations. The diagonal components $\Lambda^{\alpha\alpha}$ are generally an order of magnitude larger than the off-diagonal components $\Lambda^{\alpha\beta}$ with $\alpha\neq\beta$.

All of the inputs to the inversion have now been defined and computed. We are now ready to discuss the inversion procedure.


\section{SOLA Inversion}
\label{inversion}

We want to estimate, for example, $u_x(\br)$  given a set of travel times $\tau^\alpha$, the kernels $\bK^\alpha$, and the noise covariance $\Lambda^{\alpha\beta}$ (Eq.~[\ref{tau-u}]). We adopt the SOLA scheme developed by Pijpers \& Thompson (1992). The strategy is to search for a set of coefficients $w^\alpha$ such that an estimate of $u_x(\br)$ is
\begin{equation}
\label{tt-weight}
\overline{u_x}(\br)  := \sum_{j,\alpha} w^\alpha(\br_j-\br) \tau^\alpha(\br_j)  \approx  u_x(\br) .
\end{equation}
 The above definition of $\overline{u_x}$ is equivalent to 
\begin{equation}
\overline{u_x}(\br)  = \sum_k \bcK(\br_k-\br) \cdot \bu(\br_k)  + \sum_{j,\alpha}w^\alpha(\br_j-\br)n^\alpha(\br_j) , \; \;
\end{equation}
where
\begin{equation}
\label{ave_kernels}
\bcK(\br) :=  \sum_{j,\alpha} w^\alpha(\br_j) \bK^\alpha(\br-\br_j)
\end{equation}
is the averaging kernel. The averaging kernel is vector valued, i.e. $\bcK = (\cK_x,\cK_y)$. Ideally the $x$-component of the averaging kernel, $\cK_x$, should be a delta function of position and $\cK_y$ should vanish in order to recover $u_x$ perfectly. This is not possible in general because of noise and a limited set of travel times. Instead, in the spirit of SOLA, we attempt to choose weights so that $\cK_x(\br)$ will resemble a prescribed target function, $\cT_x(\br)$, which is localized around $\br={\bf 0}$.  In practice, we take a Gaussian function for $\cT_x(\br)$ with dispersion $\sigma$, such that 
\begin{equation}
\label{target}
\bcT(\br) =  \bxhat \frac{e^{-r^2/2\sigma^2}}{2\pi\sigma^2} .
\end{equation}
where $r=\|\br\|$ and $\| \cdot \|$ is the 2D vector norm. The total integral of $\cT_x(\br)$ is one. If we can construct an averaging kernel $\bcK \approx \bcT$ with $\sigma$ small, then $\overline{u_x}(\br)$ will be an estimate of $u_x(\br')$ in the neighborhood $\|\br'-\br\| < \sigma$.  As a result we call $\sigma$, or FWHM$\equiv2\sigma\sqrt{2\ln 2}$, the target resolution.  The effective resolution depends on how well the averaging kernel matches the target function.

In general, OLA inversions seek a balance between the resolution of the solution and the error magnification that propagates through the inversion due to the uncertainties of the measurement. To quantify this more precisely, we define two measures. The first measure gives the misfit between the averaging kernel and the chosen target function,
\begin{equation}
\label{misfit}
{\rm misfit}^2 = \sum_{i} \left\|\bcK(\br_i) - \bcT(\br_i)\right\|^2 .
\end{equation}
One wants the misfit to be as small as possible. The second measure quantifies the error in $\overline{u_x}$:
\begin{equation}
\label{error}
{\rm error}^2 = \sum_{i,j, \alpha,\beta} w^\alpha(\br_i)\Lambda^{\alpha\beta}(\br_i-\br_j) w^{\beta}(\br_j) .
\end{equation}
 It is also desirable to keep this term as small as possible. The balance of the two measures defined in equations~(\ref{misfit}) and (\ref{error}) is achieved by minimizing
\begin{equation}
\label{minimize}
{\rm misfit}^2 + \mu \,{\rm error}^2,
\end{equation}
with respect to $w^\alpha$, where $\mu$ is a trade-off, or regularization parameter. The minimization of expression~(\ref{minimize}) is also subject to the constraint
\begin{equation}
\label{constraints}
\sum_i\bcK(\br_i)=  \bxhat .
\end{equation}
The trade-off parameter is varied through a range of values. The  value for $\mu$ is chosen so that one is content with the amount of misfit and the amount of error that results from the inversion.

Upon carrying out the minimization of expression~(\ref{minimize}) with respect to the weights, one finds that for all indices $k$ and measurement types $\gamma$, 
\begin{eqnarray}
\label{full-inv}
\sum_{j,\alpha} A_{k\gamma, j \alpha} w^{\alpha}(\br_j)  = b_{k\gamma} ,
\end{eqnarray}
where we have defined
\begin{eqnarray}
A_{k\gamma, j \alpha} &:=& \sum_i \bK^\alpha(\br_i-\br_j)\cdot\bK^{\gamma}(\br_i-\br_k)  \label{eq.the2terms}
\nonumber \\  &&
 +\,\mu \Lambda^{\gamma\alpha}(\br_k-\br_j) , \\
b_{k\gamma} &:=&\sum_{i} \bK^\gamma(\br_i-\br_k)\cdot\bcT(\br_i) .  
\end{eqnarray}
In addition, the constraint~(\ref{constraints}) implies
\begin{equation}
\sum_{j,\alpha} \left[ \sum_i \bK^\alpha(\br_i-\br_j) \right] w^\alpha(\br_j) = \bxhat
\label{eq.cons}
\end{equation}
Taken together, equations~(\ref{full-inv}) and~(\ref{eq.cons}) lead to a system of linear equations that can be written as
\begin{equation}
\label{mat-inv}
\bA \bw= \bb .
\end{equation}
A computational advantage of the SOLA formulation is that the target function is contained in $\bb$, and so the matrix $\bA$ need not be inverted for each new target that satisfies the same constraint \citep{pijpers1992}. 

There are inherent problems in the solution given by $\bA^{-1}\bb$, since the matrix $\bA$ may be ill-conditioned.  The matrix $\bA$ may have small singular values which result in oscillatory singular vectors: upon inversion of the matrix, the high-frequency components are amplified and the corresponding solution tends to be meaningless. This is a commonly occurring problem in inversions. To rectify this, we regularize the two matrices corresponding to the two terms in equation~(\ref{eq.the2terms}) by performing a singular value decomposition (SVD) and removing the smallest singular values, a process referred to as truncated SVD. This technique is commonly used in helioseismic inversions \citep[e.g.,][]{jcd1993}.


\section{Results}
\label{results}

\begin{figure}[t]
   \includegraphics[width=3.5 in, clip]{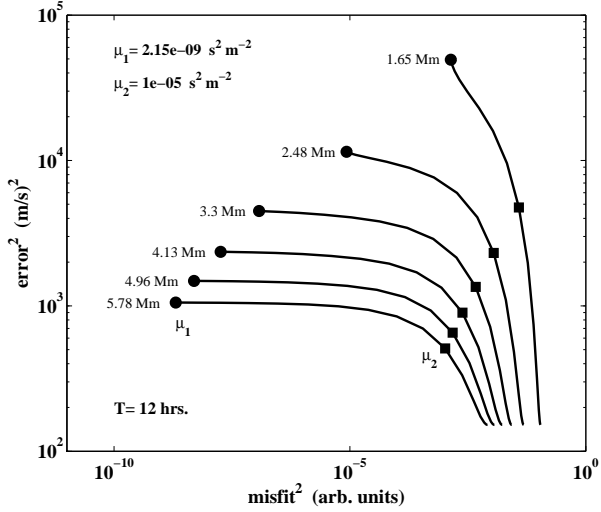}
\centering
   \caption{A set of L-curves from the inversion. Each curve represents a different FWHM of the target function, noted at the top left of the curve, and each point corresponds to a different value of the regularization parameter $\mu$. The circles and squares denote where $\mu_1$ and $\mu_2$ lie on each curve, respectively.}
   \label{fig:l_curves}
\end{figure}

We solve equation~(\ref{mat-inv}) for many sets of inversion coefficients     $w^\alpha$ using a range of values for the trade-off parameter $\mu$. This is done for different FWHM values of the target function as well. The results  can be studied in error-misfit space by computing the 'L-curves' \citep{hansen1998},   shown in Figure~\ref{fig:l_curves}.

\begin{figure}[t]
   \includegraphics[width=3.5 in, clip]{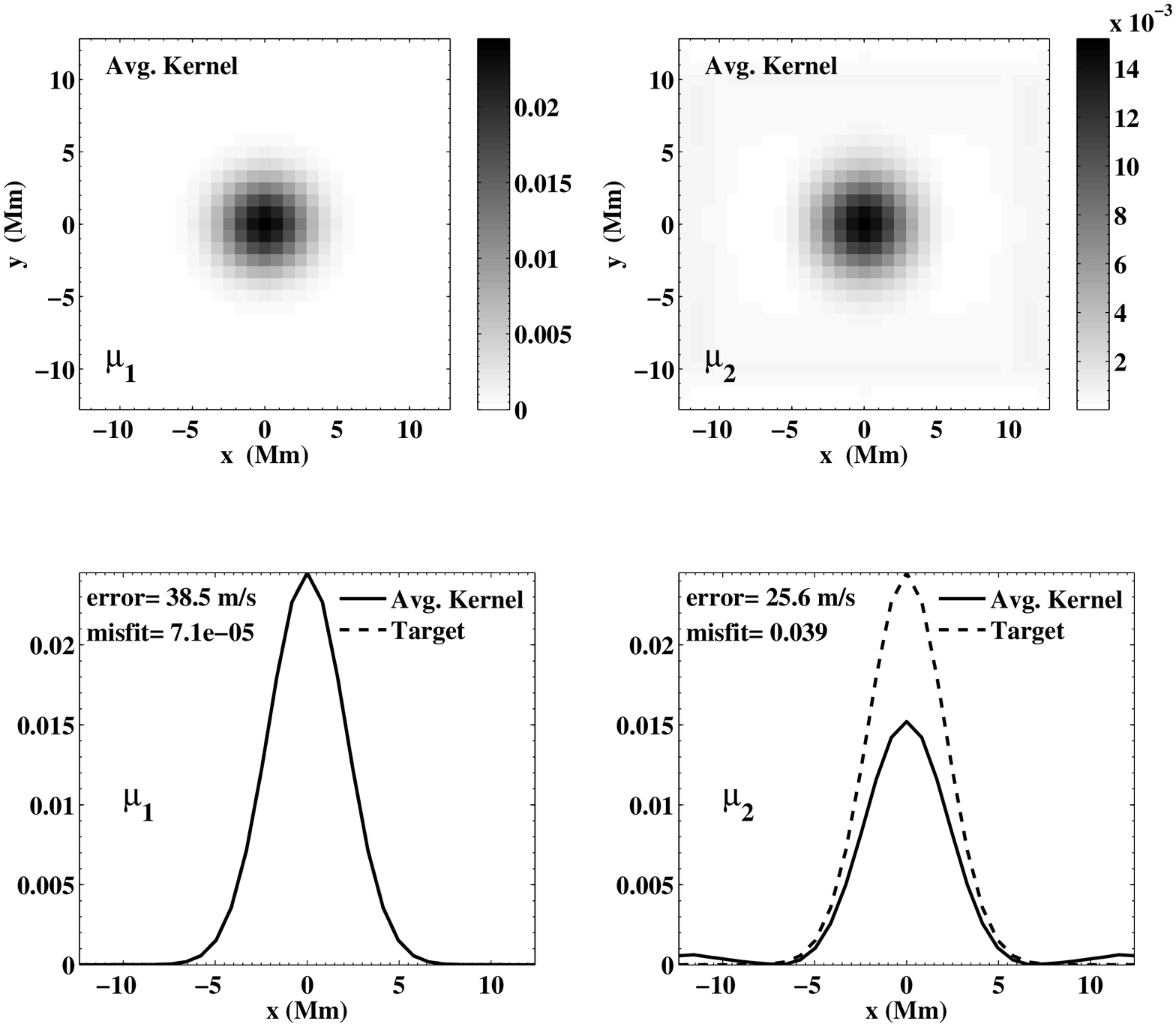}
\centering
   \caption{Averaging kernels $\cK_x$ for two different values of $\mu$, at fixed
     FWHM$=4.96$~Mm. The 2 panels on the left are for $\mu_1$, while the two on the
     right are for $\mu_2$. The top row is the averaging kernel for each $\mu$. The bottom row shows a cut along the $y=0$
     line for each averaging kernel (solid line), as well as a cut through the
     target function (dashed line). The value of the error and misfit is also given for each $\mu$. The spatial sum of each  $\cK_x$ is 1. $T$=12~hrs. Note that $\cK_y=0$.}
   \label{fig:misfits}
\end{figure}

Each curve in Figure~\ref{fig:l_curves} corresponds to a particular value of the FWHM, and each point on each curve corresponds to a different set of weights. In this case, there are 45 values of $\mu$ which span over 15 orders of magnitude to produce each L-curve. The goal is to choose a particular set of weights so that one is satisfied with the trade-off between the misfit and error.

Highlighted in Figure~\ref{fig:l_curves} are two specific values of the regularization parameter, $\mu=\mu_1$ and $\mu=\mu_2$, which represent two possible trade-off choices.  The points on the  L-curves given by $\mu=\mu_2$ are  ones chosen by an algorithm that finds the `corner' of discreet L-curves for general inverse problems  \citep[see][]{hansen2006}. The `corner' is generally considered to be the point of the best trade-off. However, in this example, choosing the weights at the points given by $\mu=\mu_1$ seems much more appropriate. For instance, consider the L-curves with the larger values of FWHM in Figure~\ref{fig:l_curves}. As one increases $\mu$ from $\mu_1$ to $\mu_2$, the increase in misfit is nearly six orders of magnitude, while the decrease in error  is only about a factor of two.

\begin{figure}[]
   \includegraphics[width=3.5 in, clip]{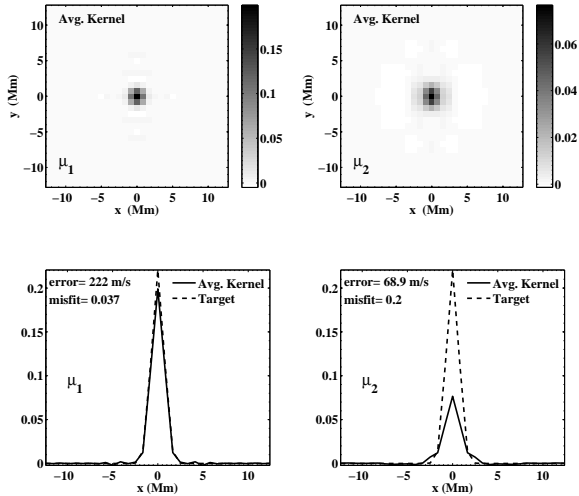}
\centering
   \caption{Same as  Figure~\ref{fig:misfits}, but for ${\rm FWHM}=1.65$~Mm.}
   \label{fig:misfits2}
\end{figure}

\begin{figure}[]
   \includegraphics[width=3.5 in, clip]{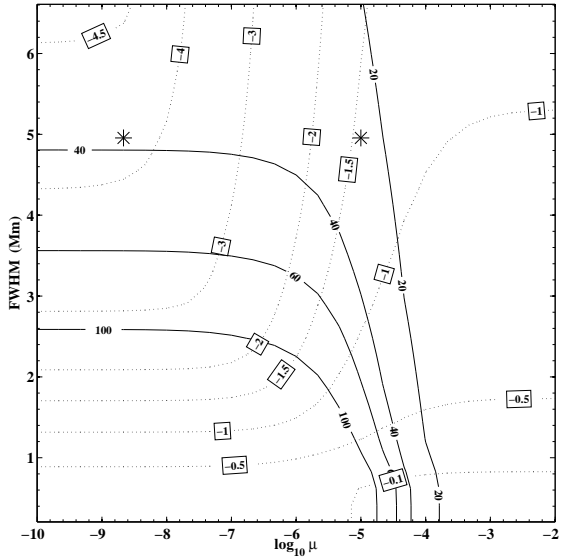}
\centering
   \caption{Contour plot of error and misfit from an inversion, as functions of FWHM and $\log_{10}\mu$. The solid lines are contours of constant error, in units of m s$^{-1}$. The dotted
     lines are  contours of constant
     $\log_{10}({\rm misfit})$, whose values are given in the boxes. The left and right stars correspond to FWHM=4.96~Mm and $\mu_1$ and $\mu_2$, respectively. As in all previous plots, $ T=12$~hrs.}
   \label{fig:err_misfit_contour}
\end{figure}

\begin{figure*}[]
   \includegraphics[width=6.8 in, clip]{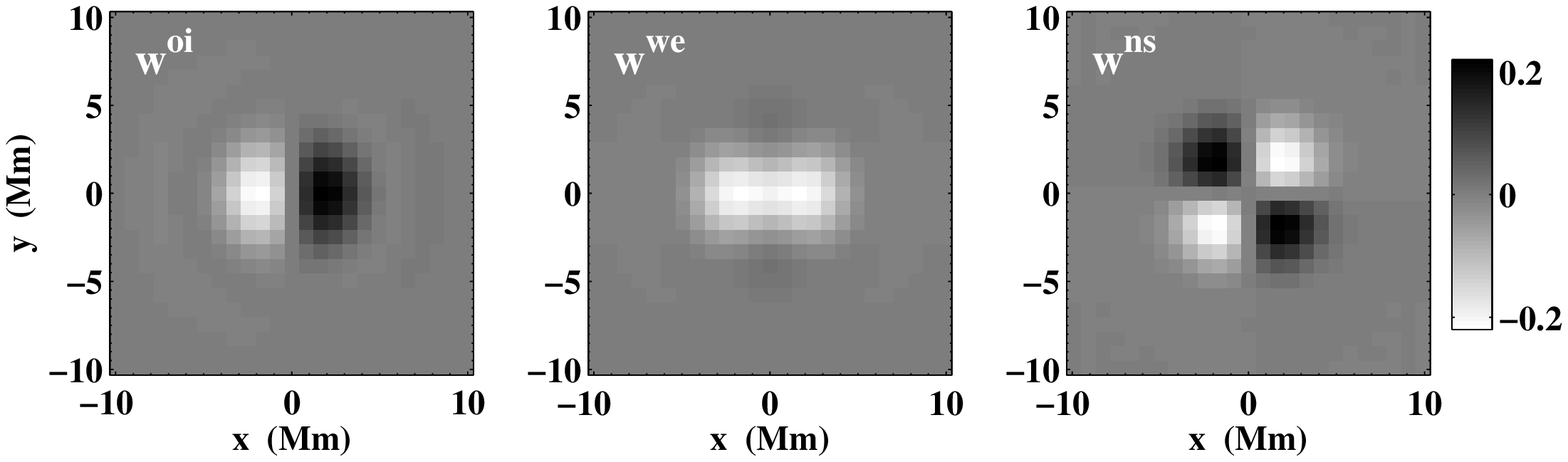}
\centering
   \caption{Inversion coefficients $w^\alpha$, corresponding to $\mu=\mu_1$, FWHM=4.96~Mm, and $T=12$~hrs. The units are (m/s) s$^{-1}$. The averaging kernel is by definition obtained by convolving these coefficients with the kernels according to equation~(\ref{ave_kernels}).}
   \label{fig:weights}
\end{figure*}

The difference between $\mu_1$ and $\mu_2$  is further explored in Figure~\ref{fig:misfits}, where we examine  the misfit more closely by plotting cuts through the averaging kernels $\cK_x$ and the target function $\cT_x$, at  $\mu_1$ and $\mu_2$ and FWHM=4.96~Mm. For this example $\cK_y=0$ according to equation~(\ref{constraints}). One sees that even though the error at $\mu=\mu_2$ is modestly lower than the error at $\mu=\mu_1$,  the averaging kernels at $\mu=\mu_2$ do not match the target function very well. In particular, there is some structure away from the central peak, which is undesirable, while the averaging kernel and target match almost exactly at $\mu=\mu_1$.  Therefore, in what follows, we choose to study the set of weights corresponding to $\mu=\mu_1$ for each FWHM. With this choice  of vanishing misfit, one may use the terms resolution and FWHM interchangeably.

We note, however, that for a very small FWHM, as in Figure~\ref{fig:misfits2}, it might be reasonable to choose a slightly larger value for $\mu$ than $\mu=\mu_1$. The L-curve for this FWHM = $1.65$~Mm (Fig.~\ref{fig:l_curves}) demonstrates that the error decreases very rapidly with very little increase in misfit as $\mu$ is increased. However, because we are using only one travel distance in this study,  we will not consider such small FWHM any further  since the error is too large. Using many travel distances would be necessary to lower the noise at such a high resolution.

Another helpful way to visualize  the trade-off is to study plots of  contours of constant error and misfit in the FWHM - $\mu$ space, as in Figure~\ref{fig:err_misfit_contour} \citep{pijpers1994}. This figure shows again that if one demands a small misfit, as in the middle to upper left region of the plot, achieving a small error requires settling for a poorer resolution. Ideally, one would like to be in the lower right part of the plot, where a small FWHM (high resolution) accompanies a small error. However, the misfit there is large and the averaging kernels are not localized at all.

The inversion coefficients $w^\alpha$ corresponding to $\mu=\mu_1$ and FWHM = $4.96$~Mm are given in Figure~\ref{fig:weights}. According to equation~(\ref{tt-weight}), the weights average the travel-time measurements to give an estimate of the solution. 

\begin{figure}[]
   \includegraphics[width=3.5 in, clip]{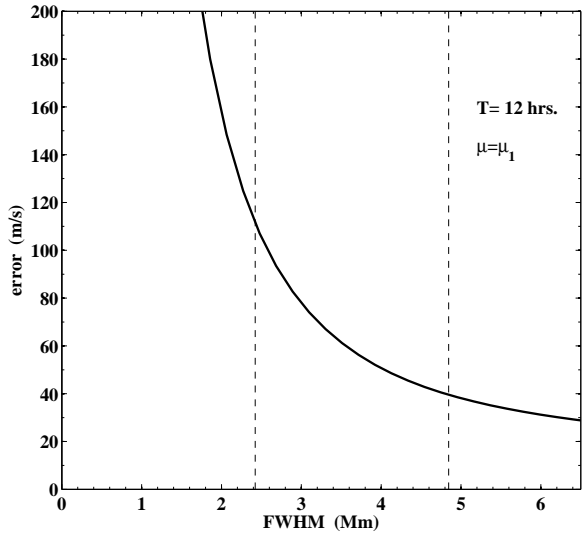}
\centering
   \caption{Error as a function of the FWHM of the target function. The two dashed vertical lines denote the value of one wavelength and one-half wavelength of f modes at 3~mHz.}
   \label{fig:error_v_width}
\end{figure}
\begin{figure}[]
   \includegraphics[width=3.5 in, clip]{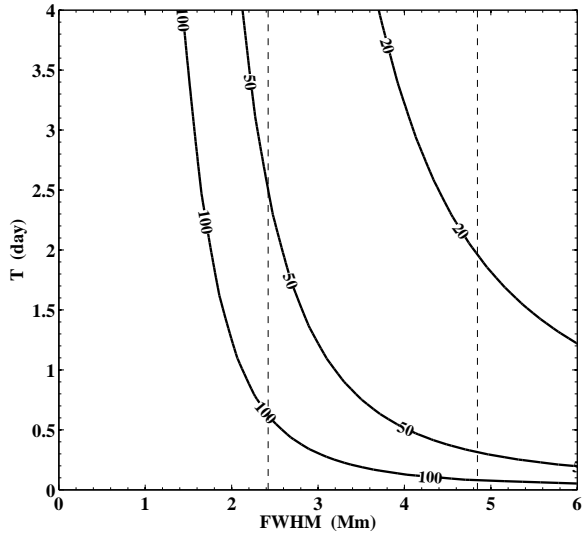}
\centering
   \caption{Contour plot of error for different observing times $T$ and  FWHM. The lines of constant error are in units of m s$^{-1}$. The error is computed using the set of inversion coefficients at $\mu=\mu_1$. The two dashed vertical lines denote the value of one wavelength and one-half wavelength of f modes at 3~mHz.}
   \label{fig:contour_err}
\end{figure}

The typical dependence of the error on the FWHM for an observation time $T$=12~hrs is shown in Figure~\ref{fig:error_v_width}. In order to achieve a reasonable error (say $<$~20 m s$^{-1}$) for $T$=12~hrs, one cannot hope for sub-wavelength resolution ($<$~5~Mm). Therefore, to increase the resolution and keep the same error estimate, it becomes necessary to observe for a longer time. As has already been mentioned, the covariance of the random error in the travel-time measurements scales as $T^{-1}$ (see Gizon \& Birch 2004). We use this result to estimate the  errors over a wide range of typical observation times, as shown in Figure~\ref{fig:contour_err}. The figure shows that to reach sub-wavelength resolution with a reasonable error, the observation time should span several days.  Conversely, Figure~\ref{fig:contour_err} is also very useful for choosing  values of $T$ and  resolution that will give an acceptable noise level.

\section{Discussion}
\label{discussion}

We have formulated and solved a two-dimensional SOLA inversion of f-mode travel times for flows. It was shown that this inversion procedure works, in that there exist averaging kernels that  match the target function very well. The random noise on the flow estimates, however, has been found to be quite large for one travel distance and typical values of the observation time, $T$. To lower the noise, it will be necessary to use many travel distances. 

We suggest that the inversion coefficients should be chosen to give essentially the smallest misfit between the target function and the averaging kernel. A useful consequence of this particular choice  is that one immediately knows the resolution. Another advantage is that travel-time measurements can be inverted separately for each annulus radius at fixed target resolution. This would allow for a straightforward averaging of the results of the inversions for the various annulus radii.

How should one know which target resolution to choose? A possible strategy is to choose first an acceptable level of random error on the flow estimates and to then choose the target resolution and observing time accordingly (see Fig.~\ref{fig:contour_err}).

We note that the present inversion procedure can presumably easily be generalized to infer flows in three dimensions (the size of the matrix to be inverted would remain the same). The generalization to other types of solar perturbations besides flows is straightforward, as long as the corresponding travel-time sensitivity kernels are available.










\end{document}